\documentclass[a4paper]{article}
\usepackage{amsmath}
\usepackage[affil-it]{authblk}

\begin{document}

\title{Energy and momentum in multiple metric theories}

\author{Idan Talshir
  \thanks{E-mail: \texttt{talshir@post.tau.ac.il}}}
\affil{School of Physics, Tel Aviv University, Ramat Aviv 69978, Israel \\
Department of Physics, Ariel University, Ariel 40700, Israel \\ }

\date{}

\maketitle

\begin{abstract}
We derive the expressions for canonical energy, momentum, and angular momentum for multiple metric theories. We prove that although the metric fields are generally interacting, the total energy is the sum of conserved energies corresponding to each metric. A positive energy theorem is given as a result. In addition, we present an Hamiltonian formalism for a subgroup of multi-metric theories.
\end{abstract}

\section{introduction}

Energy, momentum and angular momentum are fundamental concepts in physics.

In the framework of special relativistic field theory, i.e. a non-dynamical Minkowski background, those conserved quantities  are defined as the space integration on the currents corresponding to invariance of the action under time and space translations and rotations. The energy and momentum defined in this way are known as ``canonical''. An angular momentum can be derived from a symmetric energy momentum tensor. The canonical energy momentum tensor is not symmetric, 
but it can be symmetrized using the ``Belinfante Procedure'' \cite{Belinfante}. There is another way for obtaining an energy momentum tensor, by varying the action with respect to the background metric, then estimate the outcome ``on shell''. The second procedure gives a symmetric tensor by definition, and can be shown to be equivalent to the canonical procedure.\\
 The concept of energy and its conservation in general relativity has been a matter of debate for a long time. It can be defined in a coordinate system which is asymptotically Minkowski (and for other background metrics, see \cite{Abbott} ), and has a non-tensorial character under general coordinate transformation. Aside from the canonical procedure , an energy momentum pseudo-tensor can be constructed from the metric field equation. There are a few known procedures for doing so (the procedures of Weinberg \cite{Weinberg1} and of Landau and Lifshitz \cite{Landau1} yield directly symmetric pseudo-tensors), and they do not give the same result, although often  equal in their whole-space integrals.\\ 
 
 Multiple metric theories are theories with more than one second-rank tensor, that, beside being a dynamical field, is used for raising and lowering indices, and for the construction of affine connection. Recently they arose as a possible procedure for modifying general relativity. Milgrom suggested ``Bimond''\cite{Milgrom1}- a theory with two metrics: One metric is coupled to matter, and so determines the motion of test particles, and the other, a ``twin metric'', is used to construct, with the first (ordinary) metric, an interaction term. That interaction term becomes dominant at low accelerations (the MOND regime), and the theory can produce flat rotation curves with less amount of dark matter than General Relativity needs in order to explain the  phenomena. \\
 It should be noted that a large set of multiple-metric theories that can be approximated by the Pauli-Fierz action suffer from Boulware-Deser ghost instabilities \cite{inconsistencies}. A sub-set of these theories that avoid the ghost problem, has been constructed \cite{RandH}, but recent developments \cite{SandA} cast doubt on the relevance of these. However, BiMOND gravity theories are outside the framework of this debate, since these theories are intended to give the MOND potential in the weak field limit (for the flat rotation curve phenomenon), and not the Newtonian potential, and therefore cannot be linearized near the double Minkowski metrics, and cannot be approximated by the Pauli-Fierz action.

 An expression for the energy of a multiple metric theory can be used for determining its stability and validity and for generating a dynamics for the system.

 \section{Total energy is separable}
 In the following we find an expression for the total energy-momentum pseudo-tensor of a multiple metric theory, and prove that it is the sum of energy-momentum pseudo-tensors; each one can be numerically determined by one metric, and formally is the same as we get from non-interacting general relativity (Einstein-Hilbert action)systems.\\
\\

Without loss of generality, we take the number of metrics to be two. The proof is valid for any number of metrics .The action is:
\begin{equation} \label{twentyfive}
S = \int {L{d^4}x = } \int {(\alpha {L_G} + \beta {{\hat L}_G} + {L_{\operatorname{int} }}){d^4}x}   
\end{equation}

where the scalar densities in the parentheses are the Einstein-Hilbert scalar density for the metric  
${g_{\mu \nu }}$,${L_G} =  - \frac{1}{{16\pi G}}R\sqrt { - g}  $, a scalar density for the twin metric ${\hat g_{\mu \nu }}$, ${\hat L_G} =  - \frac{1}{{16\pi G}}\hat R\sqrt { - \hat g}  $ , and an interaction term which depends on both metric and other fields and their derivatives, i.e., $
{L_{\operatorname{int} }}[{g_{\mu \nu }};{g_{\mu \nu ,\rho }};{\hat g_{\mu \nu }};{g_{\mu \nu ,\rho }};\psi ;{\psi _{,\rho }}] $, where $\psi $ symbolizes any other fields with any tensorial properties. The coefficients $\alpha $ and $\beta$ are any constants.

 The total energy momentum pseudo-tensor for the system is the conserved current we get from invariance with respect to translation. This is the canonical energy momentum pseudo-tensor
\begin{equation} \label{twentysix}
\Theta _\beta ^\alpha  =  - L\delta _\beta ^\alpha  + {g_{lm,\alpha }}\frac{{\partial L}}{{\partial {g_{lm,\beta }}}} + {{\hat g}_{lm,\alpha }}\frac{{\partial L}}{{\partial {{\hat g}_{lm,\beta }}}} + {\psi _{,\alpha }}\frac{{\partial L}}{{\partial {\psi _{,\beta }}}}
\end{equation}
Separating the Lagrangian density contributions, we write this as
\begin{equation} \label{three}
\begin{gathered}
  \Theta _\beta ^\alpha  = \alpha \left( { - {L_G}\delta _\beta ^\alpha  + {g_{lm,\alpha }}\frac{{\partial {L_G}}}{{\partial {g_{lm,\beta }}}}} \right) + \beta \left( { - {{\hat L}_G}\delta _\beta ^\alpha  + {{\hat g}_{lm,\alpha }}\frac{{\partial {{\hat L}_G}}}{{\partial {{\hat g}_{lm,\beta }}}}} \right) +  \hfill \\
    + \left( { - {L_{\operatorname{int} }}\delta _\beta ^\alpha  + {g_{lm,\alpha }}\frac{{\partial {L_{\operatorname{int} }}}}{{\partial {g_{lm,\beta }}}} + {{\hat g}_{lm,\alpha }}\frac{{\partial {L_{\operatorname{int} }}}}{{\partial {{\hat g}_{lm,\beta }}}} + {\psi _{,\alpha }}\frac{{\partial {L_{\operatorname{int} }}}}{{\partial {\psi _{,\beta }}}}} \right) \hfill \\ 
 \end{gathered} 
\end{equation}
The first two terms on the right side of the equation are recognized as the gravity and twin gravity energy momentum pseudo tensors. Each one of them functionally depends on one and only one metric field, and its form does not depend on the interaction. It corresponds to the nonlinear part (with respect to the deviations of the field from the Minkowski metric) of the Einstein tensor. Now we would like to show that the energy momentum pseudo-vector that corresponds to the third term is equal to the sum of contributions from the energy-momentum source terms for each metric field equation. 
\subsection{Interaction term is separable}
We shall now prove that:                            
\begin{equation} \label{twentyseven}
\int {\left( { - {L_{\operatorname{int} }}\delta _\alpha ^0 + {g_{lm,\alpha }}\frac{{\partial {L_{\operatorname{int} }}}}{{\partial {g_{lm,0}}}} + {{\hat g}_{lm,\alpha }}\frac{{\partial {L_{\operatorname{int} }}}}{{\partial {{\hat g}_{lm,0}}}} + {\psi _{,\alpha }}\frac{{\partial {L_{\operatorname{int} }}}}{{\partial {\psi _{,0}}}}} \right)} {d^3}x = \smallint (\sqrt { - g} T_\alpha ^0 + \sqrt { - \hat g} \hat T_\alpha ^0){d^3}x
\end{equation}
where
\[T_\alpha ^\beta  \equiv {g_{\alpha \nu }}\frac{2}{{\sqrt { - g} }}\frac{{\delta {S_{\operatorname{int} }}}}{{\delta {g_{\nu \beta }}}};\hat T_\alpha ^\beta  \equiv {\hat g_{\alpha \nu }}\frac{2}{{\sqrt { - \hat g} }}\frac{{\delta {S_{\operatorname{int} }}}}{{\delta {{\hat g}_{\nu \beta }}}}\]
We use Ohanian's proof \cite{Ohanian} for equality of volume integrals of energy momentum tensor and the canonical energy momentum for the matter field, and we generalize it for interaction with more than one metric field.\\
The variation of the scalar density $ {L_{\operatorname{int} }} $  , under an infinitesimal coordinates transformation $\delta {x^\mu } = {\xi ^\mu }(x)$ , is:
\begin{equation} \label{twentyeight}
\delta {L_{\operatorname{int} }} =  - {({L_{\operatorname{int} }}{\xi ^\mu })_{,\mu }}{\text{ }}
\end{equation}

On the other hand, the variation as a functional of metric and metric variations is:
\begin{equation} \label{twentynine}
\begin{gathered}
  \delta {L_{\operatorname{int} }} = \frac{{\partial {L_{\operatorname{int} }}}}{{\partial {g_{a\nu }}}}\delta {g_{a\nu }} + \frac{{\partial {L_{\operatorname{int} }}}}{{\partial {{\hat g}_{a\nu }}}}\delta {{\hat g}_{a\nu }} + \frac{{\partial {L_{\operatorname{int} }}}}{{\partial {g_{a\nu ,\mu }}}}\delta {g_{\mu \nu ,\mu }} + \frac{{\partial {L_{\operatorname{int} }}}}{{\partial {{\hat g}_{a\nu ,\mu }}}}\delta {{\hat g}_{a\nu ,\mu }} + \frac{{\partial {L_{\operatorname{int} }}}}{{\partial \psi }}\delta \psi  + \frac{{\partial {L_{\operatorname{int} }}}}{{\partial {\psi _{,\mu }}}}\delta {\psi _{,\mu }} =  \hfill \\
   \hfill \\
   = \frac{{\delta {L_{\operatorname{int} }}}}{{\delta {g_{a\nu }}}}\delta {g_{a\nu }} + \frac{{\delta {L_{\operatorname{int} }}}}{{\delta {{\hat g}_{a\nu }}}}\delta {{\hat g}_{a\nu }} + \frac{{\delta {L_{\operatorname{int} }}}}{{\delta \psi }}\delta \psi  + {\left( {\frac{{\partial {L_{\operatorname{int} }}}}{{\partial {g_{a\nu ,\mu }}}}\delta {g_{a\nu }} + \frac{{\partial {L_{\operatorname{int} }}}}{{\partial {{\hat g}_{a\nu ,\mu }}}}\delta {{\hat g}_{a\nu }} + \frac{{\partial {L_{\operatorname{int} }}}}{{\partial {\psi _{,\mu }}}}\delta \psi } \right)_{,\mu }} \hfill \\ 
\end{gathered} 
\end{equation}
From \eqref{twentyeight},\eqref{twentynine} and the equation of motion $\frac{{\delta {L_{\operatorname{int} }}}}{{\delta \psi }} = 0$, we get
\begin{equation} \label{thirty}
{\left( { - {L_{\operatorname{int} }}{\xi ^\mu } - \frac{{\partial {L_{\operatorname{int} }}}}{{\partial {g_{a\nu ,\mu }}}}\delta {g_{a\nu }} - \frac{{\partial {L_{\operatorname{int} }}}}{{\partial {{\hat g}_{a\nu ,\mu }}}}\delta {{\hat g}_{a\nu }} - \frac{{\partial {L_{\operatorname{int} }}}}{{\partial {\psi _{,\mu }}}}\delta \psi } \right)_{,\mu }} = \frac{{\delta {L_{\operatorname{int} }}}}{{\delta {g_{a\nu }}}}\delta {g_{a\nu }} + \frac{{\delta {L_{\operatorname{int} }}}}{{\delta {{\hat g}_{a\nu }}}}\delta {{\hat g}_{a\nu }}
\end{equation}
First, from the right hand side of  \eqref{thirty} we obtain:
\begin{equation} \label{thiryone}
\begin{gathered}
  \frac{{\delta {L_{\operatorname{int} }}}}{{\delta {g_{a\nu }}}}\delta {g_{a\nu }} + \frac{{\delta {L_{\operatorname{int} }}}}{{\delta {{\hat g}_{a\nu }}}}\delta {{\hat g}_{a\nu }} =  \hfill \\
   = \frac{{\delta {L_{\operatorname{int} }}}}{{\delta {g_{a\nu }}}}( - {g_{a\alpha }}\xi _{,\nu }^\alpha  - {g_{\alpha \nu }}\xi _{,a}^\alpha  - {g_{a\nu ,\alpha }}{\xi ^\alpha }) + \frac{{\delta {L_{\operatorname{int} }}}}{{\delta {{\hat g}_{a\nu }}}}( - {{\hat g}_{a\alpha }}\xi _{,\nu }^\alpha  - {{\hat g}_{\alpha \nu }}\xi _{,a}^\alpha  - {{\hat g}_{a\nu ,\alpha }}{\xi ^\alpha }) =  \hfill \\
   = {(\sqrt { - g} {T^{a\nu }}{g_{a\alpha }}{\xi ^\alpha })_{,\nu }} - \sqrt { - g} T_{;\nu }^{a\nu }{g_{a\alpha }}{\xi ^\alpha } + {(\sqrt { - \hat g} {{\hat T}^{a\nu }}{{\hat g}_{a\alpha }}{\xi ^\alpha })_{,\nu }} - \sqrt { - \hat g} \hat T_{:\nu }^{a\nu }{{\hat g}_{a\alpha }}{\xi ^\alpha } \hfill \\ 
\end{gathered} 
\end{equation}
where the ";" and ":" indicate  covariant derivatives with respect to the metric and twin metric respectively. However
 
\begin{equation} \label{thirtytwo}
\sqrt { - g} T_{;\nu }^{a\nu }{g_{a\alpha }} + \sqrt { - \hat g} \hat T_{:\nu }^{a\nu }{\hat g_{a\alpha }} = \sqrt { - g} T_{\alpha ;\nu }^\nu  + \sqrt { - \hat g} \hat T_{\alpha :\nu }^\nu  = 0
\end{equation}
The above equation is true because from a variation of the interaction action, a scalar, under coordinate transformation $x{'^\mu } = {x^\mu } + {\xi ^\mu }$, we get:  
\begin{eqnarray} \label{thirtythree}
\begin{gathered}
  \delta {S_{\operatorname{int} }} = \int {{d^4}x} \left( {\frac{{\delta {L_{\operatorname{int} }}}}{{\delta {g_{a\nu }}}}\delta {g_{a\nu }} + \frac{{\delta {L_{\operatorname{int} }}}}{{\delta {{\hat g}_{a\nu }}}}\delta {{\hat g}_{a\nu }} + \frac{{\partial {L_{\operatorname{int} }}}}{{\delta \psi }}\delta \psi } \right) =  \hfill \\
   = \int {{d^4}x\tfrac{1}{2}\left( {\sqrt { - g} {T_{\mu \nu }}\delta {g^{\mu \nu }} + \sqrt { - \hat g} {{\hat T}_{\mu \nu }}\delta {{\hat g}^{\mu \nu }} + 0} \right)}  = 0 \hfill \\ 
\end{gathered} 
\end{eqnarray}
 (raising and lowering indices for variations in a specific metric, carried out using the same metric). Substituting $\delta {g^{\mu \nu }} = {\xi ^{\mu ;\nu }} + {\xi ^{\nu ;\mu }} $ and $\delta {\hat g^{\mu \nu }} = {\xi ^{\mu :\nu }} + {\xi ^{\nu :\mu }}$, and making use of the symmetry of ${{T_{\mu \nu }}}$ and ${{{\hat T}_{\mu \nu }}}$, we have:
\begin{equation} \label{thirtyfour}
\begin{gathered}
  \delta {S_{\operatorname{int} }} = \int {{d^4}x\left( {\sqrt { - g} {T_{\mu \nu }}{\xi ^{\mu ;\nu }} + \sqrt { - \hat g} {{\hat T}_{\mu \nu }}{\xi ^{\mu :\nu }}} \right)}  = 0 \hfill \\
   = \int {{d^4}x\left( {\sqrt { - g} \left( {{{(T_\mu ^\nu {\xi ^\mu })}_{;\nu }} - T_{\mu ;\nu }^\nu {\xi ^\mu }} \right) + \sqrt { - \hat g} \left( {{{(\hat T_\mu ^\nu {\xi ^\mu })}_{:\nu }} - \hat T_{\mu :\nu }^\nu {\xi ^\mu }} \right)} \right)}  \hfill \\ 
\end{gathered} 
\end{equation}
The first and third addends in the integrand can be written in the form
\[{(T_\mu ^\nu {\xi ^\mu })_{;\nu }}\sqrt { - g}  = \frac{\partial }{{\partial {x^\nu }}}\left( {\sqrt { - g} T_\mu ^\nu {\xi ^\mu }} \right)\,\,\,\,\,\,{(\hat T_\mu ^\nu {\xi ^\mu })_{:\nu }}\sqrt { - \hat g}  = \frac{\partial }{{\partial {x^\nu }}}\left( {\sqrt { - \hat g} \hat T_\mu ^\nu {\xi ^\mu }} \right)\]
and drop out as surface integrals. . The variations ${\xi ^\mu }$ are arbitrary, and we therefore obtain \eqref{thirtytwo}.    \\
The right side of \eqref{thirty} is now in the form:
\begin{equation} \label{thirtyfive}
{((\sqrt { - g} {T^{a\nu }}{g_{a\alpha }} + \sqrt { - \hat g} {\hat T^{a\nu }}{\hat g_{a\alpha }}+){\xi ^\alpha })_{,\nu }}
\end{equation}
For the variations of the metrics in the left side of \eqref{thirty} we insert
 $\delta {g_{lm}} =  - {g_{lm,\mu }}{\xi ^\mu } - {g_{l\mu }}\xi _{,m}^\mu  - {g_{lm}}\xi _{,\mu }^\mu $ and $\delta {{\hat g}_{lm}} =  - {{\hat g}_{lm,\mu }}{\xi ^\mu } - {{\hat g}_{l\mu }}\xi _{,m}^\mu  - {{\hat g}_{lm}}\xi _{,\mu }^\mu $, and for the variation of the field $\psi $ we use the more general expression:
 \[\delta {\psi _A} =  - {\psi _{A,\alpha }}{\xi ^\alpha } - {\Lambda _{AB\alpha }}^\beta \xi _{,\beta }^\alpha {\psi _B}\]
 where capital letters A,B stands for all the tensorial indexes of the field $\psi $ and ${\Lambda _{AB\alpha }}^\beta $ is a constant matrix, the exact form depends only on the tensorial properties of $\psi $. We get:
\begin{equation} \label{twoone}
\begin{gathered}
  {\left( { - {L_{\operatorname{int} }}{\xi ^\mu } - \frac{{\partial {L_{\operatorname{int} }}}}{{\partial {g_{lm,\mu }}}}\left( { - {g_{lm,\rho }}{\xi ^\rho } - {g_{l\rho }}\xi _{,m}^\rho  - {g_{lm}}\xi _{,\rho }^\rho } \right) - \frac{{\partial {L_{\operatorname{int} }}}}{{\partial {{\hat g}_{lm,\mu }}}}\left( { - {{\hat g}_{lm,\rho }}{\xi ^\rho } - {{\hat g}_{l\rho }}\xi _{,m}^\rho  - {{\hat g}_{lm}}\xi _{,\rho }^\rho } \right)} \right)_{,\mu }} +  \hfill \\
   + {\left( {\frac{{\partial {L_{\operatorname{int} }}}}{{\partial {\psi _{A,\beta }}}}\left( {{\psi _{A,\alpha }}{\xi ^\alpha } + {\Lambda _{AB\alpha }}^\beta \xi _{,\beta }^\alpha {\psi _B}} \right)} \right)_{,\mu }} = {\left( {\left( {\sqrt { - g} {T^{a\nu }}{g_{a\alpha }} + \sqrt { - \hat g} {{\hat T}^{a\nu }}{{\hat g}_{a\alpha }}} \right){\xi ^\alpha }} \right)_{,\nu }} \hfill \\ 
\end{gathered}  
\end{equation}

Collecting coefficients of first derivatives of $ {\xi ^\mu } $ from \eqref{twoone} , we get the equation:
\begin{equation} \label{twotwo}
\begin{gathered}
   - {L_{\operatorname{int} }}\delta _\alpha ^\beta  + \frac{{\partial {L_{\operatorname{int} }}}}{{\partial {g_{lm,\beta }}}}{g_{lm,\alpha }} + \frac{{\partial {L_{\operatorname{int} }}}}{{\partial {{\hat g}_{lm,\beta }}}}{{\hat g}_{lm,\alpha }} + {\psi _{A,\alpha }}\frac{{\partial {L_{\operatorname{int} }}}}{{\partial {\psi _{A,\beta }}}} \hfill \\
   + {\left( {\frac{{\partial {L_{\operatorname{int} }}}}{{\partial {g_{l\beta ,\mu }}}}{g_{l\alpha }} + \frac{{\partial {L_{\operatorname{int} }}}}{{\partial {{\hat g}_{l\beta ,\mu }}}}{{\hat g}_{l\alpha }} + \left( {\frac{{\partial {L_{\operatorname{int} }}}}{{\partial {g_{lm,\mu }}}}{g_{lm}} + \frac{{\partial {L_{\operatorname{int} }}}}{{\partial {{\hat g}_{lm,\mu }}}}{{\hat g}_{lm}}} \right)\delta _\alpha ^\beta  + \frac{{\partial {L_{\operatorname{int} }}}}{{\partial {\psi _{A,\mu }}}}{\Lambda _{AB\alpha }}^\beta \xi _{,\beta }^\alpha {\psi _B}} \right)_{,\mu }} =  \hfill \\
   = \sqrt { - g} {T^{a\beta }}{g_{a\alpha }} + \sqrt { - \hat g} {{\hat T}^{a\beta }}{{\hat g}_{a\alpha }} \hfill \\ 
\end{gathered} 
\end{equation}
We would like to show that the third addend in the left hand side of \eqref{twotwo} is a divergence of an anti-symmetric tensor. For this we collect the addends with the second derivatives of  $ {\xi ^\mu } $ from \eqref{twoone}. The equation is:
\begin{equation} \label{twothree}
\left( {\frac{{\partial {L_{\operatorname{int} }}}}{{\partial {g_{l\beta ,\mu }}}}{g_{l\alpha }} + \frac{{\partial {L_{\operatorname{int} }}}}{{\partial {{\hat g}_{l\beta ,\mu }}}}{{\hat g}_{l\alpha }} + \left( {\frac{{\partial {L_{\operatorname{int} }}}}{{\partial {g_{lm,\mu }}}}{g_{lm}} + \frac{{\partial {L_{\operatorname{int} }}}}{{\partial {{\hat g}_{lm,\mu }}}}{{\hat g}_{lm}}} \right)\delta _\alpha ^\beta  + \frac{{\partial {L_{\operatorname{int} }}}}{{\partial {\psi _{A,\mu }}}}{\Lambda _{AB\alpha }}^\beta \xi _{,\beta }^\alpha {\psi _B}} \right)\xi _{,\beta ,\mu }^\alpha  = 0
\end{equation}
Since $\xi _{,\beta ,\mu }^\alpha$ is symmetric in $\beta$ and  $\mu$, its coefficient should be anti-symmetric in the same indices to satisfy this equation. Therefore,
\[{\left( {\frac{{\partial {L_{\operatorname{int} }}}}{{\partial {g_{l\beta ,\mu }}}}{g_{l\alpha }} + \frac{{\partial {L_{\operatorname{int} }}}}{{\partial {{\hat g}_{l\beta ,\mu }}}}{{\hat g}_{l\alpha }} + \left( {\frac{{\partial {L_{\operatorname{int} }}}}{{\partial {g_{lm,\mu }}}}{g_{lm}} + \frac{{\partial {L_{\operatorname{int} }}}}{{\partial {{\hat g}_{lm,\mu }}}}{{\hat g}_{lm}}} \right)\delta _\alpha ^\beta  + \frac{{\partial {L_{\operatorname{int} }}}}{{\partial {\psi _{A,\mu }}}}{\Lambda _{AB\alpha }}^\beta \xi _{,\beta }^\alpha {\psi _B}} \right)_{,\mu }} = W_{\alpha ,\mu }^{[\beta \mu ]}\]
Inserting this equation in \eqref{twotwo}, we get:
\begin{equation}
- {L_{\operatorname{int} }}\delta _\alpha ^\beta  + {g_{lm,\alpha }}\frac{{\partial {L_{\operatorname{int} }}}}{{\partial {g_{lm,\beta }}}} + {{\hat g}_{lm,\alpha }}\frac{{\partial {L_{\operatorname{int} }}}}{{\partial {{\hat g}_{lm,\beta }}}} + W_{\alpha ,\rho }^{[\beta \rho ]} = \sqrt { - g} {T^{a\beta }}{g_{a\alpha }} + \sqrt { - \hat g} {{\hat T}^{a\beta }}{{\hat g}_{a\alpha }}
\end{equation}
That means that the expression on the left side is different from the expression on the right side by a divergence of an anti-symmetric (pseudo-)tensor, and three-space integration on both sides would give the same result. Statement \eqref{twentyseven} is therefore proven.   
\subsection{Dependence on metrics is separable}

Now, we want to clarify the fact that although the contribution to the energy momentum from the interaction term for each field equation seems to be functionally dependent on both metrics, then, from the field equations, we can see that they can be brought to a form which depends on only one metric. The right hand side of the equations
\[\frac{{\delta {S_{\operatorname{int} }}}}{{\delta {g_{\mu \nu }}}} =  - \alpha \frac{{\delta {S_G}}}{{\delta {g_{\mu \nu }}}}\;\;\;{\mkern 1mu} {\mkern 1mu} {\mkern 1mu} {\mkern 1mu} {\mkern 1mu} {\mkern 1mu} {\mkern 1mu} {\mkern 1mu} {\mkern 1mu} {\mkern 1mu} {\mkern 1mu} {\mkern 1mu} {\mkern 1mu} {\mkern 1mu} \frac{{\delta {S_{\operatorname{int} }}}}{{\delta {{\hat g}_{\mu \nu }}}} =  - \beta \frac{{\delta {{\hat S}_G}}}{{\delta {{\hat g}_{\mu \nu }}}}\]
is functionally dependent on only one metric, so the left side is determined numerically by only one metric.
 The contribution from the gravity term (first two addends in \eqref{three}) is manifestly dependent on one and the same metric, so the total energy momentum contribution from each field equation depends on only one metric. 
More specifically, we've seen that the (non-symmetric) energy momentum pseudo tensor is equivalent to (up to space integration)
\[\begin{gathered}
  \Theta _\nu ^\mu  =  \hfill \\
  \left( {\alpha \left( { - {L_G}\delta _\nu ^\mu  + {g_{lm,\nu }}\frac{{\partial {L_G}}}{{\partial {g_{lm,\mu }}}}} \right) + 2{g_{\alpha \nu }}\frac{{\delta {S_{\operatorname{int} }}}}{{\delta {g_{\alpha \mu }}}}} \right) + \left( {\beta \left( { - {{\hat L}_G}\delta _\nu ^\mu  + {{\hat g}_{lm,\nu }}\frac{{\partial {{\hat L}_G}}}{{\partial {{\hat g}_{lm,\mu }}}}} \right) + 2{{\hat g}_{\alpha \nu }}\frac{{\delta {S_{\operatorname{int} }}}}{{\delta {{\hat g}_{\alpha \mu }}}}} \right) \hfill \\ 
\end{gathered} \]

Inserting 
\[\frac{{\delta {S_{\operatorname{int} }}}}{{\delta {g_{\alpha \mu }}}} = \frac{\alpha }{{32\pi G}}\sqrt { - g} ({R^{a\mu }} - \tfrac{1}{2}{g^{\alpha \mu }}){\mkern 1mu} {\mkern 1mu} {\mkern 1mu} {\mkern 1mu} {\mkern 1mu} {\mkern 1mu} {\mkern 1mu} \frac{{\delta {S_{\operatorname{int} }}}}{{\delta {{\hat g}_{\alpha \mu }}}} = \frac{\beta }{{32\pi G}}\sqrt { - \hat g} ({\hat R^{a\mu }} - \tfrac{1}{2}{\hat g^{\alpha \mu }})\]
               
We get (up to three-space integration)
\begin{equation} \label{seventeen}
\Theta _\nu ^\mu  = \frac{1}{{16\pi G}}\frac{\partial }{{\partial {x^\beta }}}\left[ \begin{gathered}
  \alpha \frac{{{g_{\mu \sigma }}}}{{\sqrt { - g} }}\frac{\partial }{{\partial {x^\alpha }}}( - g)({g^{\nu \sigma }}{g^{\alpha \beta }} - {g^{\nu \alpha }}{g^{\sigma \beta }}) +  \hfill \\
  \beta \frac{{{{\hat g}_{\mu \sigma }}}}{{\sqrt { - \hat g} }}\frac{\partial }{{\partial {x^\alpha }}}( - \hat g)({{\hat g}^{\nu \sigma }}{{\hat g}^{\alpha \beta }} - {{\hat g}^{\nu \alpha }}{{\hat g}^{\sigma \beta }}) \hfill \\ 
\end{gathered}  \right]
\end{equation}

Each one of the two addends is manifestly divergence free, because they are anti-symmetric in $\beta$  and $\nu$  , so there are two conserved energy momentum (pseudo)vectors, each one numerically determined  by one metric alone. \\
In fact, since we proved that the total energy-momentum pseudo-tensor is a sum of contributions that we get from general relativity actions for each metric, we can use a known result for the symmetric energy-momentum pseudo-tensor of general relativity, derived using the Belinfante procedure \cite{Bak}.\\
The symmetric energy-momentum pseudo-tensor for the multiple-metric system is then:
\begin{equation} \label{eightteen}
{\Theta ^{\mu \nu }} = \frac{1}{{16\pi G}}\frac{\partial }{{\partial {x^\alpha }}}\frac{\partial }{{\partial {x^\beta }}}\left[ \begin{gathered}
  \alpha \sqrt { - g} ({\eta ^{\mu \nu }}{g^{\alpha \beta }} - {\eta ^{\alpha \nu }}{g^{\mu \beta }} + {\eta ^{\alpha \beta }}{g^{\mu \nu }} - {\eta ^{\mu \beta }}{g^{\alpha \nu }}) +  \hfill \\
  \beta \sqrt { - \hat g} ({\eta ^{\mu \nu }}{{\hat g}^{\alpha \beta }} - {\eta ^{\alpha \nu }}{{\hat g}^{\mu \beta }} + {\eta ^{\alpha \beta }}{{\hat g}^{\mu \nu }} - {\eta ^{\mu \beta }}{{\hat g}^{\alpha \nu }}) \hfill \\ 
\end{gathered}  \right]
\end{equation}  
With this symmetric pseudo-tensor the angular momentum of the multiple metric system can be well defined.

\subsection{Boundary conditions}

The energy momentum pseudo-tensor $ {\Theta ^{\mu \nu }} $ and the energy-momentum pseudo-vector $   {P^\mu } = \int {{\Theta ^{\mu 0}}} {d^3}x $  are Lorentz-covariant, i.e. are tensor and vector under Lorentz transformations of the coordinate system. In order to insure finiteness and meaning of the quantity $ {P^\mu } $ we demand asymptotic behaviour from the interaction energy momentum tensors
\[T_\mu ^\nu  = {\rm O}({r^{ - 4}})\,\,\,\,\,\,\,\,\,\,\hat T_\mu ^\nu  = {\rm O}({r^{ - 4}})\] (in general relativity one can often assume an isolated system of masses so $T_\mu ^\nu  = 0$ outside a bounded region of space), and we assume asymptotic behaviour for the metric fields as  $ r \to \infty $:
\[{g_{\mu \nu }} \simeq {\eta _{\mu \nu }} + {\rm O}({r^{ - 1}})\,\,\,\,,\,\,\,\,{\hat g_{\mu \nu }} \simeq {\eta _{\mu \nu }} + {\rm O}({r^{ - 1}})\]
 so the gravity and twin gravity energy momentum pseudo tensors, which are quadratic in the derivatives of the metrics (first two terms in \eqref{three}, resulting from the Einstein scalars, $G$ and $\hat{G}$), go like ${r^{ - 4}}$.\\
The total energy momentum tensor, then, has the asymptotic behaviour $ {\Theta ^{\mu \nu }} = {\rm O}({r^{ - 4}}) $ so its integral over three-space converges, and has the same numeric value for every infinite spacelike hypersurface. \\ 
The energy momentum $ {P^\mu} $ is also invariant under any coordinate transformation that tends to the identity at infinity, and preserving the boundary conditions above. This is because the energy momentum pseudo tensor can be written as a divergence of a third rank pseudo tensor, as we can see from \eqref{seventeen} and \eqref{eightteen}. Taking, for example, the expression from \eqref{eightteen}, we can write this (pseudo-) tensor as a sum of anti-symmetric (pseudo-) tensors:
\[\begin{gathered}
  {\Theta ^{\mu \nu }} = \frac{1}{{16\pi G}}\left( {\frac{\partial }{{\partial {x^\alpha }}}\frac{\partial }{{\partial {x^\beta }}}\left[ \begin{gathered}
  \alpha \sqrt { - g} ({\eta ^{\mu \nu }}{g^{\alpha \beta }} - {\eta ^{\alpha \nu }}{g^{\mu \beta }}) +  \hfill \\
  \beta \sqrt { - \hat g} ({\eta ^{\mu \nu }}{{\hat g}^{\alpha \beta }} - {\eta ^{\alpha \nu }}{{\hat g}^{\mu \beta }}) \hfill \\ 
\end{gathered}  \right] + \frac{\partial }{{\partial {x^\beta }}}\frac{\partial }{{\partial {x^\alpha }}}\left[ \begin{gathered}
  \alpha \sqrt { - g(} {\eta ^{\alpha \beta }}{g^{\mu \nu }} - {\eta ^{\mu \beta }}{g^{\alpha \nu }}) +  \hfill \\
  \beta \sqrt { - \hat g} ({\eta ^{\alpha \beta }}{{\hat g}^{\mu \nu }} - {\eta ^{\mu \beta }}{{\hat g}^{\alpha \nu }}) \hfill \\ 
\end{gathered}  \right]} \right) =  \hfill \\
   = \frac{1}{{16\pi G}}\left( {\frac{\partial }{{\partial {x^\alpha }}}{A^{\alpha \mu \nu }} + \frac{\partial }{{\partial {x^\beta }}}{B^{\beta \mu \nu }}} \right) \hfill \\ 
\end{gathered} \]
 where $ {A^{\alpha \mu \nu }} =  - {A^{\mu \alpha \nu }} $ and ${B^{\beta \mu \nu }} =  - {B^{\mu \beta \nu }}$ so its whole space integral reduces to an ordinary surface integral at infinity, which is not changed by the above coordinate transformation, i.e.  
 \[\begin{gathered}
   \int {\left( {\frac{\partial }{{\partial {x^\alpha }}}{A^{\alpha 0\nu }} + \frac{\partial }{{\partial {x^\beta }}}{B^{\beta 0\nu }}} \right)} {d^3}x = \int {{{\left( {\frac{\partial }{{\partial {x^i}}}{A^{i0\nu }} + \frac{\partial }{{\partial {x^i}}}{B^{i0\nu }}} \right)}_{i = 1,2,3}}} {d^3}x \hfill \\
    \hfill \\
    = \int {\left( {{A^{i0\nu }} + {B^{i0\nu }}} \right)d{S_i}^{(2)}}  \hfill \\ 
 \end{gathered} \]
 Specifically, we get for the total energy and total momentum
 \begin{equation} \label{twothreeone}
P_{total}^\nu  = {P^\nu } + {{\hat P}^\nu }
\end{equation}
where
\begin{equation} \label{twothreetwo}
{P^0} \equiv \frac{\alpha }{{16\pi G}}\int {\left( {\frac{{\partial {g_{ij}}}}{{\partial {x^j}}} - \frac{{\partial {g_{jj}}}}{{\partial {x^i}}}} \right)} d{s^i}{\mkern 1mu} {\mkern 1mu} {\mkern 1mu} {\mkern 1mu} {\mkern 1mu} {\mkern 1mu} {\mkern 1mu} {\mkern 1mu} {\mkern 1mu} {\mkern 1mu} {{\hat P}^0} \equiv \frac{\beta }{{16\pi G}}\int {\left( {\frac{{\partial {{\hat g}_{ij}}}}{{\partial {x^j}}} - \frac{{\partial {{\hat g}_{jj}}}}{{\partial {x^i}}}} \right)} d{s^i}
\end{equation}
and
\begin{equation} \label{twothreethree}
{P^j} \equiv \frac{\alpha }{{16\pi G}}\int {\left( {\frac{{\partial {g_{kk}}}}{{\partial {x^0}}}\frac{{\partial {g_{k0}}}}{{\partial {x^k}}}{\delta _{ij}} + \frac{{\partial {g_{j0}}}}{{\partial {x^i}}} - \frac{{\partial {g_{ij}}}}{{\partial {x^0}}}} \right)d{s^i}{\mkern 1mu} {\mkern 1mu} {\mkern 1mu} {\mkern 1mu} {\mkern 1mu} {\mkern 1mu} {\mkern 1mu} {\mkern 1mu} } {{\hat P}^j} \equiv \frac{\beta }{{16\pi G}}\int {\left( {\frac{{\partial {{\hat g}_{kk}}}}{{\partial {x^0}}}\frac{{\partial {{\hat g}_{k0}}}}{{\partial {x^k}}}{\delta _{ij}} + \frac{{\partial {{\hat g}_{j0}}}}{{\partial {x^i}}} - \frac{{\partial {{\hat g}_{ij}}}}{{\partial {x^0}}}} \right)d{s^i}{\mkern 1mu} {\mkern 1mu} {\mkern 1mu} {\mkern 1mu} {\mkern 1mu} {\mkern 1mu} {\mkern 1mu} {\mkern 1mu} } 
\end{equation}

\section{Applications- positive energy theorem}

In the following we prove that the total energy of a multi-metric system is not negative under specific conditions. in the following we assume that the coefficients are positive, and we'll discuss later the other cases. This proof is a generalization of Witten's elegant proof (\cite{Witten}, \cite{Fadeev}) of the positive energy theorem for general relativity.
Defining the tetrads \[{g^{\mu \nu }} = e_a^\mu {\eta ^{ab}}e_b^\nu \,\,\,\,\,\,\,\,\,\,\,\,\,\,{\hat g^{\mu \nu }} = \hat e_a^\mu {\eta ^{ab}}\hat e_b^\nu \,\]
(indices $a,b,c$ in the tetrads are local)
,and imposing the conditions \[e_0^0 = 1;\,\,\,\,\,\,\,\,{e^{00}} = {e_{00}} =  - 1;\,\,\,\,\,\,e_0^\alpha  = e_0^i = 0\].\\
($i = 1,2,3$),
connection coefficients \[\begin{gathered}
  {\omega _{\mu ,ab}} = \tfrac{1}{2}e_\mu ^c({\Omega _{cab}} - {\Omega _{abc}} - {\Omega _{bca}})\,\,\,\,\,\,\,\,\,\,\,\,\,{\Omega _{cab}} = {e_{\nu c}}(e_a^\mu {\partial _\mu }e_b^\nu  - e_b^\mu {\partial _\mu }e_a^\nu ) \hfill \\
  {{\hat \omega }_{\mu ,ab}} = \tfrac{1}{2}\hat e_\mu ^c({{\hat \Omega }_{cab}} - {{\hat \Omega }_{abc}} - {{\hat \Omega }_{bca}})\,\,\,\,\,\,\,\,\,\,\,\,\,{{\hat \Omega }_{cab}} = {{\hat e}_{\nu c}}(\hat e_a^\mu {\partial _\mu }\hat e_b^\nu  - \hat e_b^\mu {\partial _\mu }\hat e_a^\nu ) \hfill \\ 
\end{gathered} \]
and the three-dimensional Dirac operators  
\[\begin{gathered}
  D = e_a^i{\gamma ^a}{\nabla _i}\,\,\,\,\,\,\,\,\hat D = \hat e_a^i{\gamma ^\alpha }{{\hat \nabla }_i} \hfill \\
   \hfill \\
  {\nabla _\mu } = {\partial _\mu } + \tfrac{1}{8}{\omega _{\mu ,ab}}[{\gamma ^a},{\gamma ^b}]\,\,\,\,\,\,\,\,{{\hat \nabla }_\mu } = {\partial _\mu } + \tfrac{1}{8}{{\hat \omega }_{\mu ,ab}}[{\gamma ^a},{\gamma ^b}] \hfill \\ 
\end{gathered} \]
Ordinary matter fields in general relativity fulfil the ``positivity condition'' for their energy momentum tensor. If the condition holds for each field equation in a multiple metric theory, then there is a unique solution for the Dirac equations \[D\psi  = 0\,\,\,\,\,\,\,\,\,\hat D\hat \psi  = 0\] with asymptotic behaviour 
\[\begin{gathered}
  \psi  = {\psi _0} + O\left( {\frac{1}{r}} \right)\,\,\,\,{\partial _\mu }\psi  = O\left( {\frac{1}{{{r^{1 + \alpha }}}}} \right)\,\,\,\,\,\alpha  > \tfrac{1}{2} \hfill \\
  \hat \psi  = {{\hat \psi }_0} + O\left( {\frac{1}{r}} \right)\,\,\,\,{\partial _\mu }\hat \psi  = O\left( {\frac{1}{{{r^{1 + \hat \alpha }}}}} \right)\,\,\,\,\,\hat \alpha  > \tfrac{1}{2} \hfill \\ 
\end{gathered} \]  for every constant spinors $ {\psi _0},{\hat \psi _0} $.\\
The following equations hold:
\begin{equation} \label{p1}
\begin{gathered}
  \int {e\left[ {\tfrac{1}{2}{\psi ^*}({T_{00}} + {T_{0k}}e_\alpha ^k{\gamma ^0}{\gamma ^\alpha })\psi  + \alpha {{({\nabla _i}\psi )}^*}({\nabla ^i}\psi )} \right]{d^3}x = } \tfrac{1}{4}(E(\psi _0^*{\psi _0}) + {P_\alpha }(\psi _0^*{\gamma ^0}{\gamma ^\alpha }{\psi _0})) \hfill \\
  \int {\hat e\left[ {\tfrac{1}{2}{{\hat \psi }^*}({{\hat T}_{00}} + {{\hat T}_{0k}}e_\alpha ^k{\gamma ^0}{\gamma ^\alpha })\hat \psi  + \beta {{({\nabla _i}\hat \psi )}^*}({\nabla ^i}\hat \psi )} \right]{d^3}x = } \tfrac{1}{4}(\hat E(\hat \psi _0^*{{\hat \psi }_0}) + {{\hat P}_\alpha }(\hat \psi _0^*{\gamma ^0}{\gamma ^\alpha }{{\hat \psi }_0})) \hfill \\ 
\end{gathered} 
\end{equation}
where ${P_\alpha } \equiv {\eta _{\alpha \beta }}{P^\alpha };{\hat P_\alpha } \equiv {\eta _{\alpha \beta }}{\hat P^\alpha }$ and $E,\hat E,{\hat P^j},{\hat P^j}$ are defined by equations \eqref{twothreetwo} and \eqref{twothreethree}.  From this set of equations we have

\begin{equation} \label{p3}
E \ge \left| {{P_\alpha }} \right|\,\,\,\,\,\,\,\,\hat E \ge \left| {{{\hat P}_\alpha }} \right|
 \end{equation}
 
 Now we use the proof that the total canonical energy is the sum of the energies that correspond to each metric, and the same for the canonical momentum (equation \eqref{twothreeone}), and conclude that the total energy is not negative and that the Lorentz vector $P_{total}^\nu $ is not space-like.\\
That the total energy is zero if both metrics are flat can be seen immediately by placing $ {\eta _{\mu \nu }} $ in \eqref{eightteen}.\\
  To prove that the total energy is zero only if all energy momentum tensors are zero and both metrics are flat, we use \eqref{p3} to note that if $E,\hat E = 0$ then $ {P_\alpha },{{\hat P}_\alpha } = 0 $.
  Placing these in \eqref {p1} we get the equations
\begin{equation} \label{p4}
{\nabla _i}\psi  = 0\,\,\,\,\,\,\,{\nabla _i}\hat \psi  = 0
\end{equation}
and
\begin{equation}  \label{p5}
{\psi ^*}({T_{00}} + {T_{0k}}e_\alpha ^k{\gamma ^0}{\gamma ^\alpha })\psi  = 0\,\,\,\,\,\,\,\,\,{{\hat \psi }^*}({{\hat T}_{00}} + {{\hat T}_{0k}}\hat e_\alpha ^k{\gamma ^0}{\gamma ^\alpha })\hat \psi  = 0\,
\end{equation}
Using \eqref{p4} we can show than the curvature tensors on some initial surface ${x^0} = 0$ are zero.
\begin{equation}  \label{p6}
{R_{ikab}} = 0\,\,\,\,\,\,\,{{\hat R}_{ikab}} = 0
\end{equation}
Using \eqref{p5} we can show that
\begin{equation} \label {p7}
{T_{00}} = 0\,\,\,\,\,\,\,{{\hat T}_{00}} = 0\,\,
\end{equation} 
From \eqref {p7} and the ``positivity condition'' we get ${T_{\mu \nu }} = 0\,\,\,\,{{\hat T}_{\mu \nu }} = 0\,$, so ${R_{\mu \nu }} = 0\,\,\,\,{{\hat R}_{\mu \nu }} = 0\,$. Using this and \eqref{p6} we get ${R_{0k0a}} = 0\,\,\,\,\,\,\,{{\hat R}_{0k0a}} = 0$, so altogether we get 
\[{R_{\mu \nu ab}} = 0\,\,\,\,\,\,\,{{\hat R}_{\mu \nu ab}} = 0\]
on the surface ${x^0} = 0$\\
We proved that the energies $E,\hat E$ are separately conserved, i.e., constant in time, so the above proof is valid for every surface ${x^0}$; the curvature tensors are then zero in all points, and both metrics are flat. This completes the proof that the total energy is positive for every non-trivial configuration in multi-metric theories.\\
The proof was constructed under the assumption that the coefficients $\alpha , \beta$ of the ``kinetic'' terms are both positive. This assumption is crucial. Moreover, if one of the coefficient is negative, say $\beta<0$, and it's corresponding interaction energy-momentum tensor ${\hat T_{\mu \nu }}$ fulfils the ``negativity condition'',so that the matrix ${\hat T_{00}} + {\hat T_{0k}}e_\alpha ^k{\gamma ^0}{\gamma ^\alpha }$ has only negative eigenvalues, then we can use the above construction to prove that the energy $\hat E$ is negative or can be made negative with some Lorentz transformation. In that case, the total energy may be positive or negative for different multi-metric field configurations. These considerations can be used to exclude theories with non-positive coefficients.

\section{Hamiltonian formalism}
In this section we present the Hamiltonian, the canonical variables and the constraint surfaces explicitly for some multimetric interactions.\\ 
The theory of constrained Hamiltonian system was introduced by Dirac \cite{Dirac1},\cite{Dirac2}, and extended by him to general relativity. One of the best known and most used Hamiltonian formalism for general relativity is the ADM or 3+1 formalism \cite{ADM}.\\
To apply the Hamiltonian formalism in relativistic field theory, we need to separate the spacetime into space + time, and to distinguish the time coordinate from the other coordinates. This separation was done in previous sections, when we chose the boundary conditions for the metric.\\ In order to find the canonical variables of a multi-metric theory, We want to follow the path shown in  \cite{FandP} the canonical variables for GR are identified. We define the conditions under which one can proceed on this route.\\
We present the gravitational parts of the action \eqref{twentyfive} in the form

\begin{equation}
\begin{gathered}
  \int {\left( {\alpha {L_G} + \beta {{\hat L}_G}} \right)} {d^4}x =  \hfill \\
   = \smallint \left( \begin{gathered}
  \frac{{  \alpha }}{{16\pi G}}\left( { - \Gamma _{\mu \rho }^\rho {\partial _\nu }\left( {\sqrt { - g} {g^{\mu \nu }}} \right) + \Gamma _{\mu \nu }^\rho {\partial _\rho }\left( {\sqrt { - g} {g^{\mu \nu }}} \right) + \sqrt { - g} {g^{\mu \nu }}\left( {\Gamma _{\mu \sigma }^\rho \Gamma _{\rho \nu }^\sigma  - \Gamma _{\mu \nu }^\rho \Gamma _{\rho \sigma }^\sigma } \right)} \right) +  \hfill \\
  \frac{{  \beta }}{{16\pi G}}\left( { - \hat \Gamma _{\mu \rho }^\rho {\partial _\nu }\left( {\sqrt { - \hat g} {{\hat g}^{\mu \nu }}} \right) + \hat \Gamma _{\mu \nu }^\rho {\partial _\rho }\left( {\sqrt { - \hat g} {{\hat g}^{\mu \nu }}} \right) + \sqrt { - \hat g} {{\hat g}^{\mu \nu }}\left( {\hat \Gamma _{\mu \sigma }^\rho \hat \Gamma _{\rho \nu }^\sigma  - \hat \Gamma _{\mu \nu }^\rho \hat \Gamma _{\rho \sigma }^\sigma } \right)} \right) \hfill \\ 
\end{gathered}  \right){d^4}x \hfill \\ 
\end{gathered} 
\end{equation}
so that the fields ${\Gamma _{\mu \nu }^\rho }$ and ${\hat \Gamma _{\mu \nu }^\rho }$ are varied independently and not as functionals of the metrics. If the interaction ${L_{\operatorname{int} }}$ depends only on the metrics on not on their derivatives, then the  equations obtained from variation of the metrics are identical to the equations of motion in the ``metric'' formalism, and the equations obtained from variations of the connection coefficients $\Gamma _{\mu \nu }^\rho  $ give the desired connection between them and the metrics as Christoffel connections. The condition that the interaction depends only on the metrics is sufficient, however not necessary, for getting the correct equations of motions, but we need it for a later argument. \\
Under the same assumption that the interaction depends only on the metrics and not on their derivatives, the constraint equations 
\begin{equation} 
\label{H2}
\frac{{\delta S}}{{\delta \Gamma _{\mu \nu }^\rho }} = \frac{{\partial L}}{{\partial \Gamma _{\mu \nu }^\rho }} = 0\,\,\,\,\,\,\,\,\,\frac{{\delta S}}{{\delta \hat \Gamma _{\mu \nu }^\rho }} = \frac{{\partial L}}{{\partial \hat \Gamma _{\mu \nu }^\rho }} = 0
\end{equation}
for $(\rho ,\mu ,\nu \,) = (0,j,k)$, contain the fields $ \Gamma _{i0}^0,\Gamma _{i0}^k,\Gamma _{ij}^k,\hat \Gamma _{i0}^0,\hat \Gamma _{i0}^k,\hat \Gamma _{ij}^k $ in a linear manner, so these fields can be expressed as a solution of \eqref{H2} with the fields $(\Gamma _{ik}^0,{h^{\mu \nu }},\hat \Gamma _{ik}^0,{{\hat h}^{\mu \nu }})$ where ${h^{\mu \nu }} \equiv \sqrt { - g} {g^{\mu \nu }}$ and ${{\hat h}^{\mu \nu }} \equiv \sqrt { - \hat g} {{\hat g}^{\mu \nu }}$, or alternatively with the variables $({\pi _{ik}},{q^{ik}},{h^{0\mu }},{{\hat \pi }_{ik}},{{\hat q}^{ik}},{{\hat h}^{0\mu }})$ where 
\begin{equation}
\begin{gathered}
  {q^{ik}} \equiv {h^{0i}}{h^{0k}} - {h^{00}}{h^{ik}},\,\,{{\hat q}^{ik}} \equiv {{\hat h}^{0i}}{{\hat h}^{0k}} - {{\hat h}^{00}}{{\hat h}^{ik}} \hfill \\
  {\pi _{ik}} \equiv  - \frac{1}{{{h^{00}}}}\Gamma _{ik}^0,\,\,{{\hat \pi }_{ik}} \equiv  - \frac{1}{{{{\hat h}^{00}}}}\hat \Gamma _{ik}^0 \hfill \\ 
\end{gathered} 
\end{equation}
The addends in the Lagrangian that contain time derivative can be collected and written in the form
\begin{equation}
\alpha {\pi _{ik}}{\partial _0}{q^{ik}} + \beta {{\hat \pi }_{ik}}{\partial _0}{{\hat q}^{ik}}
\end{equation}
That is, in a form of a kinetic term with canonical variables.\\
If ${T_{0\mu }} \equiv \frac{2}{{\sqrt { - g} }}\frac{{\delta {S_{\operatorname{int} }}}}{{\delta {g^{0\mu }}}}$ and ${\hat T_{0\mu }} \equiv \frac{2}{{\sqrt { - \hat g} }}\frac{{\delta {S_{\operatorname{int} }}}}{{\delta {{\hat g}^{0\mu }}}}$ can be written as functionals of $({q^{ik}},{{\hat q}^{ik}})$, then, under the assumption of double-Minkowski boundary conditions (and therefore dropping total 3-divergence terms), it is possible to put the rest of the Lagrangian  
in the form
\begin{equation}
- {H_{Tot}} - \left( {\frac{1}{{{h^{00}}}} + 1} \right){C_0} - \left( {\frac{{{h^{0k}}}}{{{h^{00}}}}} \right){C_k} - \left( {\frac{1}{{{{\hat h}^{00}}}} + 1} \right){{\hat C}_0} - \left( {\frac{{{{\hat h}^{0k}}}}{{{{\hat h}^{00}}}}} \right){{\hat C}_k}
\end{equation}
with

\begin{equation}
\begin{gathered}
  {H_{Tot}} = H + \hat H \hfill \\
  H = -{C_0} - \frac{\alpha }{{16\pi G}}{\partial _i}{\partial _k}{q^{ik}}{\mkern 1mu} {\mkern 1mu} {\mkern 1mu} ;{\mkern 1mu} {\mkern 1mu} {\mkern 1mu} \hat H = -{{\hat C}_0} - \frac{\alpha }{{16\pi G}}{\partial _i}{\partial _k}{{\hat q}^{ik}} \hfill \\ 
\end{gathered} 
\end{equation}
and
\begin{equation}
\begin{gathered}
  {C_0} = \frac{\alpha }{{16\pi G}}\left( {{q^{ik}}{q^{mn}}\left( {{\pi _{ik}}{\pi _{mn}} - {\pi _{ik}}{\pi _{mn}}} \right) + \gamma {R_3}} \right) - {T_{00}}{\mkern 1mu} {\mkern 1mu} {\mkern 1mu} {\mkern 1mu} {{\hat C}_0} = \frac{\beta }{{16\pi G}}\left( {{{\hat q}^{ik}}{{\hat q}^{mn}}\left( {{{\hat \pi }_{ik}}{{\hat \pi }_{mn}} - {{\hat \pi }_{ik}}{{\hat \pi }_{mn}}} \right) + \hat \gamma {{\hat R}_3}} \right) - {{\hat T}_{00}} \hfill \\
  {C_k} = \frac{\alpha }{{16\pi G}}\left( {2{\nabla _k}\left( {{q^{il}}{\pi _{il}}} \right) - 2{\nabla _l}\left( {{q^{il}}{\pi _{ik}}} \right)} \right) - {T_{0k}}{\mkern 1mu} {\mkern 1mu} {\mkern 1mu} {\mkern 1mu} {\mkern 1mu} {{\hat C}_k} = \frac{\beta }{{16\pi G}}\left( {2{{\hat \nabla }_k}\left( {{{\hat q}^{il}}{{\hat \pi }_{il}}} \right) - 2{{\hat \nabla }_l}\left( {{{\hat q}^{il}}{{\hat \pi }_{ik}}} \right)} \right) - {{\hat T}_{0k}} \hfill \\ 
\end{gathered} 
\end{equation}
where ${R_3},{\hat R_3}$ are the curvatures of the sub-metrics on the surfaces ${x^0} = const$ and are expressed with the canonical variables and not with derivatives, in the  same way that the four dimensional curvature scalars are expressed with metrics and connections. In this form, one can see that $ {H_{Tot}} $ is actually the generalized Hamiltonian of the system, $ \left( {{C_0},{C_k},{{\hat C}_0},{{\hat C}_k}} \right)  $ define constraint surfaces, and their coefficients that are composed from the fields $ {h^{0\mu }} $ and $ {{\hat h}^{0\mu }} $, are Lagrange multipliers.\\
In this case the number of degrees of freedom is the number of metric fields multiplies  the number of degrees of freedom of general relativity (not including matter and other non-metric fields). In the general case, however, this is not necessarily true, even if the interaction depends only on the metrics. In general relativity, Bianchi identities reduce the number of independent fields (metric terms). In multi-metric theories, Bianchi identities are still valid, but the identities themselves are not independent, so the number of the fields which are constraines by these identities is smaller than the number of metrics times four Bianchi identities. Another way to look on this issue is to by counting independent variations on the surfaces ${x^0} = const$. We can, with a suitable coordinate transformation, make any change of the value of the ``velocities'' ${g_{\mu 0,0}}$ without changing the value of the fields ${g_{\mu \nu }}$ and the velocities ${g_{ik,0}}$, but the same  transformation determines the change in ${{\hat g}_{\mu 0,0}}$. \\
We see that the energy obtained, which is actually the numerical value of the Hamiltonian on the constraint surfaces, can be presented as a sum of two expressions, each one depending functionally on one metric. This is a result already proved in a previous section. It should be emphasized, however, that we proved the result for much more general case than the one discussed in this chapter.\\
 It is true in principal, that if  we get the form of the generalized Hamiltonian and the constraints, one can calculate the energy. But finding this form, in the more general case where the interaction depends also on metric derivatives, may be a difficult task, and is a matter for a further research.

\section{Summary and discussion}
We saw that energy, momentum and angular momentum of a multi-metric theory with interaction can be well defined and divided into independently conserved quantities, where each one depends numerically on one metric. With appropriate boundary conditions, total energy and momentum can be presented as surface integrals. Although we have in general, only one diffeomorphism invariance, the required boundary conditions are kept under general Poincare transformation of the coordinated. These results were applied in a  generalization of positive energy theorem, and we established a criterion for excluding multi-metric theories with non-positive coefficients, based on the theorem's proof. In the last section we presented an Hamiltonian formalism for multi-metric theories with interactions which does not involve metric derivatives, and identified the canonical variables and constraint surfaces.\\
The multi-metric theory, like Einstein's general relativity, is constrained, and its naive Hamiltonian formulation, with Legendre transform and the total energy expression, must be modified to take into account these constraints. This can be done by adding to the total energy expression some combination of the constraints, with coefficients which are determined by the algorithm of Dirac \cite{Dirac1},\cite{Dirac2}.
The ADM Hamiltonian formalism \cite{ADM} or an equivalent one \cite{fadeev} may be used  to build  an  Hamiltonian for some multi-metric system, based on the proof that multi-metric system energy  is the simple sum of the canonical energies for general relativity, as defined by the ADM expression.\\
The formalism that is presented in this paper can, in principal, be used for elimination of bi(multi) metric theories by experiment. For example, if the interaction term in  specific instant is zero (the interaction term does not depends explicitly in space-time coordinates, but the fields are) we can see if there is an ``energy leak''. Since we have shown energy conservation for each metric, non-zero interaction is a necessary and sufficient condition for a change in the energy flux of the gravity, which is a measurable quantity. That is, if there is no change in the gravity energy-flux, there is no multi-metric interaction. This is not a trivial result: If energy conservation were only for the whole Lagrangian system, then there would have been the possibility that a change in the interaction energy is approximately equal in magnitude and with the opposite sign of the change in the energy of the other metric fields, so the gravity energy would still be conserved.\\

\subsection*{Acknowledgements}
I want to thank Lawrence P. Horwitz and Marcelo Schiffer for useful comments, directions, and interesting discussions.
I gratefully acknowledge financial support from Ariel University Center.
I want to thank Stanley Deser  who referred me to recent developments on the research of multi-metric theories, and for comments that helped me to locate issues which needed to be clarified. 
I want to thank the anonymous referees (Physical Review D) for constructive criticism.

\end{document}